\documentclass{article}
\usepackage{graphicx} 
\usepackage[margin=1in]{geometry}

\usepackage{booktabs}
\usepackage{natbib} 
\usepackage{xcolor}
\usepackage{ulem}
\usepackage{hyperref}
\usepackage{amsmath}
\usepackage[utf8]{inputenc} 
\usepackage{textcomp} 
\usepackage{float}
\usepackage[affil-it]{authblk}
\usepackage[most]{tcolorbox}   
\usepackage{subcaption}        
\usepackage{caption}
\usepackage{tcolorbox}
\usepackage{amssymb}
\usepackage{graphicx}          
\tcbset{sharp corners=all, colback=white, colframe=black!15, boxrule=0.6pt}

\author[1,2]{Yazdan Babazadeh Maghsoodlo\thanks{Corresponding author: \texttt{ybabazad@uwaterloo.ca}}}

\author[2]{Madhur Anand}
\author[1]{Chris T. Bauch}
\affil[1]{Department of Applied Mathematics, University of Waterloo, Waterloo, Ontario, Canada,}
\affil[2]{School of Environmental Sciences, University of Guelph, Guelph, Ontario, Canada}

\title{When Simple is Enough: Binary Models Capture Social Complexity in Coupled Human–Environment Systems.}

\date{September 2025}

\begin{document}

\maketitle

\section{Abstract}

Models of coupled human--environment systems often face a tradeoff between realism and tractability. 
Spectrum opinion models, where social preferences vary continuously, offer descriptive richness but are computationally demanding and parameter-heavy. 
Binary formulations, in contrast, are analytically simpler but raise concerns about whether they can capture key socio--ecological feedbacks. 
Here we systematically compare binary and spectrum social models across four benchmark settings: (i) replicator dynamics coupled to a climate--carbon system, (ii) FJ opinion dynamics coupled to the climate--carbon system, (iii) replicator dynamics coupled to a forest--grassland ecological system, and (iv) FJ opinion dynamics coupled to a forest--grassland ecological system. We employ the relative integrated absolute error (RIAE) to quantify deviations between binary ($N=2$) and spectrum ($N=100$) formulations of social opinion dynamics in their feedback with ecological subsystems. 
Across systematic parameter sweeps of learning rates, reluctance, conformity, susceptibility, runaway amplitudes, and ecological turnover, the binary formulation typically tracks its spectrum counterpart to within \(\leq 15\%\) for most parameter combinations. Deviations beyond this threshold arise primarily under very high social susceptibility or near-vanishing ecological turnover, where additional opinion modes and nonlinear feedbacks become consequential. We therefore present the binary formulation as a practical surrogate, not a universal replacement. As a rule of thumb, it is adequate when susceptibility is moderate, ecological turnover is appreciable, and runaway amplitudes are not extreme; in high-susceptibility or low-turnover regimes (especially near critical transitions), the full-spectrum model is preferable. This framing is intended to guide readers on when a binary reduction is sufficient versus when full-spectrum detail is warranted.

\section{Introduction}

Human societies and ecosystems are tightly intertwined, with feedbacks between social behavior and ecological processes shaping the trajectory of global environmental change \cite{lenton2008tipping,levin1998ecosystems,folke2011reconnecting,nyborg2016social,liu2007complexity,binder2013comparison,lade2013regime}. Collective decisions about land use, resource exploitation, and climate mitigation can accelerate or prevent ecological tipping points, while ecological changes in turn reshape social preferences and behaviors \cite{innes2013impact,henderson2016alternative,babazadeh2025social,scheffer2001catastrophic,farahbakhsh2023drivers}. In recent years, studies of coupled socio-ecological models have grown rapidly, reflecting a recognition that integrating human behavior into ecological modeling is essential for understanding resilience, forecasting critical transitions, and designing effective policy responses\cite{farahbakhsh2022modelling,farahbakhsh2024tipping,sigdel2019convergence,bury2019charting,beckage2022incorporating,shin2025climate,beckage2020earth,homayounfar2021coupled,schlueter2012new,phillips2020compound}.

A central challenge in socio-ecological modeling lies in how to represent social behavior. Some models treat opinions or strategies as continuous variables, allowing individuals to occupy positions along a spectrum that captures nuanced attitudes and gradual shifts in preference \cite{noah1999social,degroot1974reaching, kumar2025opinion,reynolds2025unique,bernardo2021achieving,anderson2019recent,zhou2024friedkin,biondi2023dynamics}. Others reduce social states to binary categories, such as cooperation versus defection, mitigation versus non-mitigation, or forest versus grassland preference \cite{bury2019charting,babazadeh2025social,innes2013impact,sigdel2017competition,farahbakhsh2024tipping,bauch2005imitation,menard2021conflicts,tavoni2012survival}. While spectrum models offer greater descriptive richness, binary models remain far more common because they are simpler to analyze, easier to parameterize, and computationally tractable \cite{farahbakhsh2024tipping}. Yet it remains unclear whether this simplification meaningfully alters system-level outcomes, or whether binary representations can faithfully approximate the dynamics of more complex spectrum models.

In this study, we address this gap by testing the robustness of binary social models against their spectrum counterparts in coupled socio-ecological systems. Specifically, we ask: Do binary models capture the essential dynamics of more complex spectrum social models? To answer this, we combine two ecological models with two distinct representations of social behavior, yielding four scenarios that allow systematic comparison across binary and spectrum formulations \cite{babazadeh2025social,kumar2025opinion,innes2013impact,bury2019charting}. This design enables us to assess whether the choice of social model meaningfully influences ecological outcomes, or whether binary models are sufficient approximations for capturing the critical feedbacks between human behavior and ecological dynamics.


Our results across four scenarios indicate that binary social models often approximate spectrum-model outcomes with practical accuracy, preserving the dominant feedbacks and system-level behaviors that govern ecological transitions. Our aim here is not to argue that binary models are universally best; rather, we provide guidance on when a binary reduction is sufficient. In many parameter regimes, such as moderate social susceptibility, appreciable ecological turnover, and away from runaway conditions, the binary formulation delivers substantial computational efficiency with little loss of explanatory power. By contrast, under high susceptibility or very slow turnover, the full spectrum model is preferable.

These findings carry important implications for socio-ecological modeling. By validating the use of binary social models, our work supports the adoption of simpler formulations in large-scale or policy-oriented contexts where computational efficiency and interpretability are paramount. Binary models are easier to parameterize, more transparent in their assumptions, and more tractable for integration into complex ecological frameworks. Demonstrating that they approximate spectrum models well also helps unify ecological and social modeling approaches, offering a common foundation for exploring resilience, tipping points, and intervention strategies across a wide range of coupled human--environment systems.

\section{Methods}

\subsection{Forest--grassland mosaic model}

As our first ecological model, we adopt the forest--grassland mosaic system developed by Innes et al \cite{innes2013impact}. In this framework, the ecosystem is composed of the proportion of land covered by forest $F(t)$ and grassland $G(t)$, where $F+G=1$. The dynamics of forest cover are governed by recruitment of new trees into grassland and loss of forest through natural disturbances. This leads to the differential equation

\begin{equation}
\frac{dF}{dt} = w(F) F (1-F) - v F,
\end{equation}

where $v$ is the rate at which forest reverts to grassland due to natural disturbances, and $w(F)$ is a recruitment function that depends on forest cover. The recruitment term $w(F)F(1-F)$ captures the density-dependent establishment of new forest: recruitment requires both existing forest (seed sources) and available grassland (space for colonization). 

The function $w(F)$ accounts for the mediating role of fire, which strongly suppresses recruitment when forest cover is sparse, but has weaker effects once dense stands of trees are established. This is represented by a sigmoidal function,

\begin{equation}
w(F) = \frac{c}{1 + e^{-k \frac{F}{1-F}+ b}} ,
\end{equation}

where $c$ controls the maximum recruitment rate, $b$ sets the baseline recruitment at low forest cover, and $k$ controls the abruptness of the transition between low and high recruitment regimes. This model yields alternative stable states: a grassland-dominated equilibrium ($F^*=0$) and an interior equilibrium ($F^*>0$) where both forest and grassland coexist, depending on parameter values and initial conditions.

The forest--grassland mosaic dynamics can be extended to include human influence through a harvesting function $J(x)$ that represents the net effect of social preferences on land conversion. The coupled ecological equation becomes

\begin{equation}
\frac{dF}{dt} = w(F) F (1-F) - v F + J(x),
\end{equation}

where $J(x)$ captures the contribution of social opinion to deforestation or reforestation.

\subsection{Climate--carbon model}

For our second ecological model, we consider a reduced Earth system model with four carbon pools and an energy balance for global-mean temperature anomaly \cite{lenton2000land,muryshev2015lag,babazadeh2025social,bury2019charting}. Let $C_a(t)$, $C_{oc}(t)$, $C_v(t)$, and $C_{so}(t)$ denote, respectively, atmospheric, oceanic, vegetation, and soil carbon anomalies relative to baseline stocks $(C_{a0}, C_{oc0}, C_{v0}, C_{so0})$. Let $T(t)$ be the temperature anomaly (in~$^\circ$C) relative to a baseline $T_0$, the dynamics are

\begin{align}
    \frac{dC_{at}}{dt} &= \epsilon(t) - P + R_{veg} + R_{so} - F_{oc} \\
    \frac{dC_{oc}}{dt} &= F_{oc} \\
    \frac{dC_{veg}}{dt} &= P - R_{veg} - L \\
    \frac{dC_{so}}{dt} &= L - R_{so} \\
    c\frac{dT}{dt} &= (F_d - \sigma T^4)a_E
\end{align}

where $\varepsilon(t)$ is anthropogenic emissions, $a_E$ is Earth’s emitting area, $c$ is the effective heat capacity of the climate system, and $\sigma$ is the Stefan–Boltzmann constant. Full details of this model, including the definitions of all functions and parameters, are provided in the Supplementary Materials.

Exogenous gross emissions $E(t)$ are reduced by mitigation behavior according to a
time-varying mean effort $\bar m(t)$ defined in the social layer. In general,
$\bar m(t)$ can be defined over different ranges depending on the formulation of
the social model. For example, if $\bar m(t)\in[-1,1]$, then effective emissions
are given by
\begin{equation}
\varepsilon(t) \;=\; E(t)\ \frac{1}{2}\big(1 - \bar m(t)\big),
\end{equation}
so that $\bar m(t)=-1$ corresponds to no mitigation ($\varepsilon=E$) and
$\bar m(t)=1$ corresponds to full mitigation ($\varepsilon=0$).

\subsection{Replicator social model}

As our first social model, we employ a replicator dynamics formulation in which the population is partitioned into $N$ belief categories \cite{babazadeh2025social,bury2019charting}. Let $x_i(t)$ denote the fraction of the population in category $i=0,\dots,N-1$, with $\sum_{i=0}^{N-1} x_i=1$. Each category is associated with a belief value
\[
m_i \;=\; -1 \;+\; \frac{2i}{N-1}, \qquad i=0,\dots,N-1,
\]
so that opinions range continuously from $-1$, indicating a strong grassland preference or low climate mitigation, to $+1$, indicating a strong forest preference or high climate mitigation.

Each category has an associated utility $U_i$ and the mean utility in the population is $\bar U = \sum_i x_i U_i(F)$. Social imitation dynamics are then given by the replicator equation,
\[
\frac{dx_i}{dt} \;=\; \kappa\, x_i \big(U_i(F) - \bar U\big),
\]
where $\kappa$ is the social learning rate, categories with above-average utility grow in prevalence, while those with below-average utility decline. The replicator social model is coupled to the ecological dynamics by including the mean belief
\[
\langle m \rangle \;=\; \sum_{i=0}^{N-1} m_i \, x_i,
\]

 This framework naturally interpolates between binary and continuous opinion models by varying $N$.  
\begin{itemize}
    \item For $N=2$, the opinion space collapses to two categories, $m_0=-1$ and $m_1=+1$. This corresponds to a binary social model, in which individuals can only prefer either grassland (equivalently, climate mitigation) or forest (equivalently, climate non-mitigation). The replicator equation then reduces to a two-state competition between these discrete preferences.
    \item For large $N$ (In this work $N=100$), the categories $m_i$ densely cover the interval $[-1,1]$, yielding a continuous spectrum of opinions. In this case the replicator dynamics approximate an opinion distribution evolving smoothly in response to ecological feedback.
\end{itemize}
In this way, the replicator formulation provides a unifying framework that can generate either binary or continuous social dynamics depending on the choice of $N$.

\subsection{FJ social model}

We model opinion dynamics with a Friedkin--Johnsen–type process on a bounded, scalar opinion $X_i(t)\in[-1,1]$ for each agent $i=1,\dots,n$. Let $x_i^0$ denote the agent's private anchor, and let $\lambda_i\in[0,1]$ be the agent's susceptibility to social influence (heterogeneous, drawn around $\lambda_0$ and clipped to $[0,1]$) \cite{kumar2025opinion,noah1999social}. Opinions evolve according to
\begin{equation}
\frac{dX_i}{dt}
\;=\;
\Psi\Big[
\underbrace{\lambda_i\Big(\sum_{j\neq i} w_{ij}(X_j - X_i) + R\Big)}_{\text{social influence + runaway forcing}}
\;+\;
\underbrace{(1-\lambda_i)(x_i^0 - X_i)}_{\text{anchoring to private belief}}
\Big],
\qquad X_i\in[-1,1],
\label{eq:fj}
\end{equation}
with reflecting/clipping at the bounds. The pairwise weights are distance–decaying in opinion space,
\begin{equation}
w_{ij} \;=\; \exp\!\Big(-\frac{|X_i-X_j|}{\text{A}}\Big), \qquad w_{ii}=0,
\end{equation}
where $A$ is the sensitivity of interaction strength to the opinion distance between individuals, and $R$ represents an exogenous runaway driver that shifts opinions in response to changes in the system and
$\Psi$ is an overall social adjustment rate.

The same dynamical system \eqref{eq:fj} produces a spectrum model when we use the raw opinions $X_i\in[-1,1]$ and average them,
\[
\bar X_{\text{spec}}(t) \;=\; \frac{1}{n}\sum_{i=1}^n X_i(t),
\]
 To obtain a binary version that retains only group polarity, we map each opinion to its sign and then average:
\[
\tilde X_i(t) \;=\; \operatorname{sgn}\big(X_i(t)\big)\in\{-1,+1\}, 
\qquad
\bar X_{\text{bin}}(t) \;=\; \frac{1}{n}\sum_{i=1}^n \tilde X_i(t).
\]

\subsection{Comparison Metric}

To evaluate whether binary and continuous social models produce comparable outcomes when
coupled to the same ecological dynamics, we require a metric that captures the difference between two
time series in a way that is both interpretable and scale-free. We employ the relative integrated
absolute error (RIAE), which is an $L^1$ absolute relative gap measure \cite{reich2016case}.

Formally, given two trajectories $f_1(t)$ and $f_2(t)$ defined on a time interval $[t_0, T]$, the RIAE is

\begin{equation}
\mathrm{RIAE}(f_1,f_2) \;=\; 
\frac{\int_{t_0}^T |f_1(t)-f_2(t)| \, dt}
     {\int_{t_0}^T |f_{\mathrm{ref}}(t)| \, dt}
\end{equation}

where $f_{\mathrm{ref}}$ is a chosen reference trajectory (typically one of the two models). The numerator measures the integrated $L^1$ distance
between the two trajectories, while the denominator normalizes by the total magnitude of the reference
series. This normalization makes the RIAE unitless and directly interpretable as a percentage
difference.

\begin{figure}[h!]
    \centering
    \includegraphics[width=\textwidth]{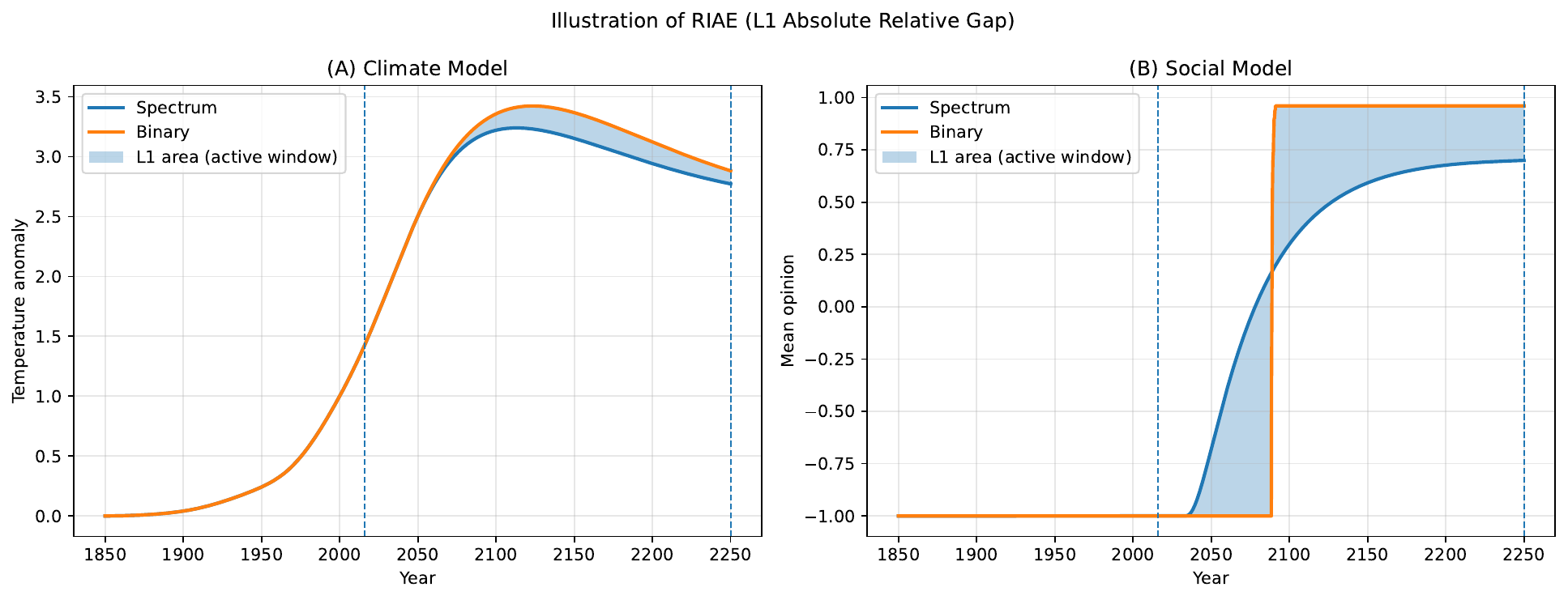}
    \caption{
    Illustration of the relative integrated absolute error (RIAE) metric. (A) Climate dynamics: comparison between spectrum and binary social models, 
    with the shaded region indicating the integrated absolute difference after the 
    onset of social dynamics ($t_0$) until the end of the evaluation window. 
    (B) Social dynamics: comparison of mean opinion trajectories under spectrum 
    and binary formulations, with shaded areas likewise representing the integrated 
    $L^1$ gap.}
    \label{fig:riae-illustration}
\end{figure}

In all simulations, the social model is initially inactive to allow the ecological variables to evolve without
social feedback. We then activate the social dynamics at time $t_0$, and only from this point do we
compute the RIAE. The integration therefore spans the interval from social onset until either (i) the system
approaches its new equilibrium state or (ii) a specified maximum evaluation time is reached, depending on
the scenario. This ensures that the comparison focuses exclusively on the time window when binary versus
continuous opinion dynamics can exert influence on the ecological model \ref{fig:riae-illustration}.

This metric has several desirable properties for our setting. First, it is sensitive to cumulative differences
across time, rather than to isolated pointwise deviations, thereby accounting for the entire ecological
trajectory over the chosen evaluation window. Second, because it is normalized, it allows comparison
across different ecological variables (e.g.\ temperature, forest cover, or mean opinion) that may have
different scales and units. Third, expressing the RIAE as a percentage facilitates interpretation: for
example, $\mathrm{RIAE}=0.10$ indicates that the binary model deviates from the spectrum model by an
average of 10\% of the reference magnitude over the evaluation period.

\section{Results}

\subsection{Coupling Between the Replicator Social Model and the Climate--Carbon Model}

The coupling between the replicator social dynamics and the climate--carbon model operates through two
feedback channels, linking temperature anomaly to social utilities and mean social opinion to emissions.

The temperature anomaly $T(t)$ enters the replicator dynamics by shaping the utility of each opinion
category. Specifically, the utility assigned to category $i$ is
\begin{equation}
U_i(T) \;=\; -\alpha_i \;+\; \beta_i \, f(T) \;+\; \delta \, x_i ,
\end{equation}
where $\alpha_i$ is a baseline reluctance parameter. The values of $\alpha_i$ are assigned along a linear
grid from $0$ up to $\alpha_{\max}$, with the number of steps equal to the number of opinion categories
$N$. $\beta_i$ is the mitigation effort associated with category $i$ that goes from 0 to $\beta_{\max}$ similar to $\alpha_i$, and $\delta$ is a conformity term. The parameter $\beta_{\max}$
controls the amplitude or strength of the perceived climate cost, scaling how strongly $f(T)$ influences
the utility of each category.

\begin{equation}
f(T) \;=\; \frac{f_{\max}}{1 + \exp[-\omega \,(T - T_c)]}
\end{equation}
increases with temperature anomaly and represents the perceived benefit of mitigation under ecological
stress. In this way, rising $T$ increases the utility of higher-mitigation categories, biasing imitation
dynamics toward stronger mitigation.

Conversely, the social state influences the climate system by reducing anthropogenic emissions. The
population distribution $x(t)$ determines the mean mitigation effort
\begin{equation}
\bar m(t) \;=\; \sum_i m_i x_i(t),
\end{equation}
which rescales gross exogenous emissions $E(t)$ to yield effective emissions,
\begin{equation}
\varepsilon(t) \;=\; E(t) \,\frac{\big(1 - \bar m(t)\big)}{2}.
\end{equation}
These effective emissions $\varepsilon(t)$ drive the atmospheric carbon equation in the climate--carbon
model.

For $E(t)$ we used empirical emissions data up to the year 2017 \cite{gilfillan2021cdiac}. Beyond this point, future emissions were
modeled with a saturating function of time \cite{babazadeh2025social,bury2019charting}:
\begin{equation}
E(t) \;=\; E(2017) \;+\; \frac{(t-2017)\,\varepsilon_{\max}}{\,t - 2017 + s\,},
\qquad t > 2017,
\end{equation}
where $\varepsilon_{\max}$ is the maximum additional emissions rate and $s$ is a saturation constant. This
functional form produces a gradual increase that asymptotically approaches a finite level, preventing
unrealistic divergence of future emissions.

To systematically compare the binary and spectrum replicator social models in the coupled climate--carbon
system, we performed parameter sweeps across four key social parameters ($\kappa$, $\alpha$, $\beta$,$\delta$). In each case, we varied one
parameter across a prescribed range in 10 steps, while keeping the others fixed at baseline values that corresponds a moderate social learning rate ($\kappa = 0.05$), a moderate maximum reluctance
parameter ($\alpha_{\max} = 1$), unit social conformity strength ($\delta = 1$), and unit maximum mitigation effort
($m_{\max}=1$). For each sweep, we considered two values of the critical temperature threshold,
$T_c \in \{2.0, 3.0\}$, representing low and high climate sensitivity scenarios.

The parameter ranges explored were:
\begin{itemize}
    \item $\kappa$: 0.01 to 0.10,
    \item $\alpha_{\max}$: 0.0 to 2.0 
    \item $\beta_{\max}$: 0.0 to 2.0
    \item $\delta$: 0.0 to 2.0 
    
\end{itemize}

\begin{figure}[htbp]
\centering
\includegraphics[width=\textwidth]{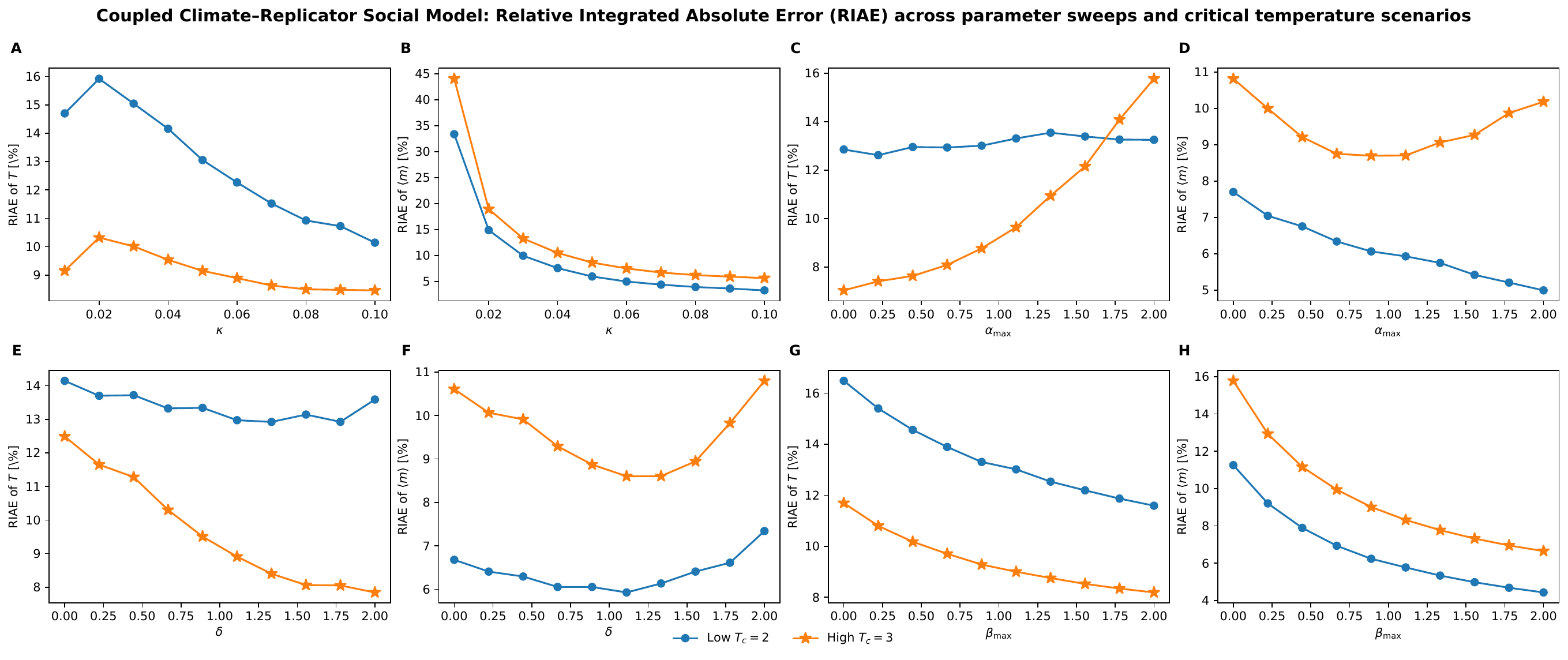}
\caption{%
Comparison of binary ($N=2$) and spectrum ($N=100$) replicator models coupled to the climate–carbon system.
Panels show RIAE of temperature anomaly $T(t)$ (left) and mean mitigation $\langle m(t)\rangle$ (right) across parameter sweeps for $\kappa$, $\alpha_{\max}$, $\delta$, and $\beta_{\max}$. 
}
\label{fig:binary_vs_spectrum1}
\end{figure}

For each parameter setting, we simulated both the binary ($N=2$) and spectrum ($N=100$) replicator models
under otherwise identical ecological conditions. The coupled model was run from the year 1800 to 2200,
with the social model kept inactive until 2017. From 2017 onward, the social dynamics were activated and
the simulation continued until 2200. We then computed the relative integrated absolute error (RIAE) between
the two social model formulations for both the climate temperature anomaly $T(t)$ and the mean mitigation
trajectory $\langle m(t)\rangle$, restricting the integral to the active period of social dynamics
(2017–2200). 
Figure~\ref{fig:binary_vs_spectrum1} shows the outcome of these sweeps, highlighting how the approximation
error between binary and spectrum models depends on the underlying social parameters and the assumed
critical temperature threshold. Overall, the results in Figure~\ref{fig:binary_vs_spectrum1} demonstrate that the deviations between binary 
and spectrum replicator models remain small across a wide range of parameter values. For nearly all 
sweeps away from the baseline settings, the relative integrated absolute error (RIAE) of both the climate 
trajectory $T(t)$ and the mean mitigation trajectory $\langle m(t)\rangle$ stays below 15\%. This indicates 
that the binary formulation closely tracks the spectrum model even under substantial variation of social 
learning rate, reluctance, conformity, and maximum mitigation effort, confirming the robustness of the 
binary approximation in the coupled climate--social system. Full information on the simulation details, including the complete table of all parameter values used, is provided in the Supplementary Information.

\begin{figure}[htbp]
\centering
\includegraphics[width=\textwidth]{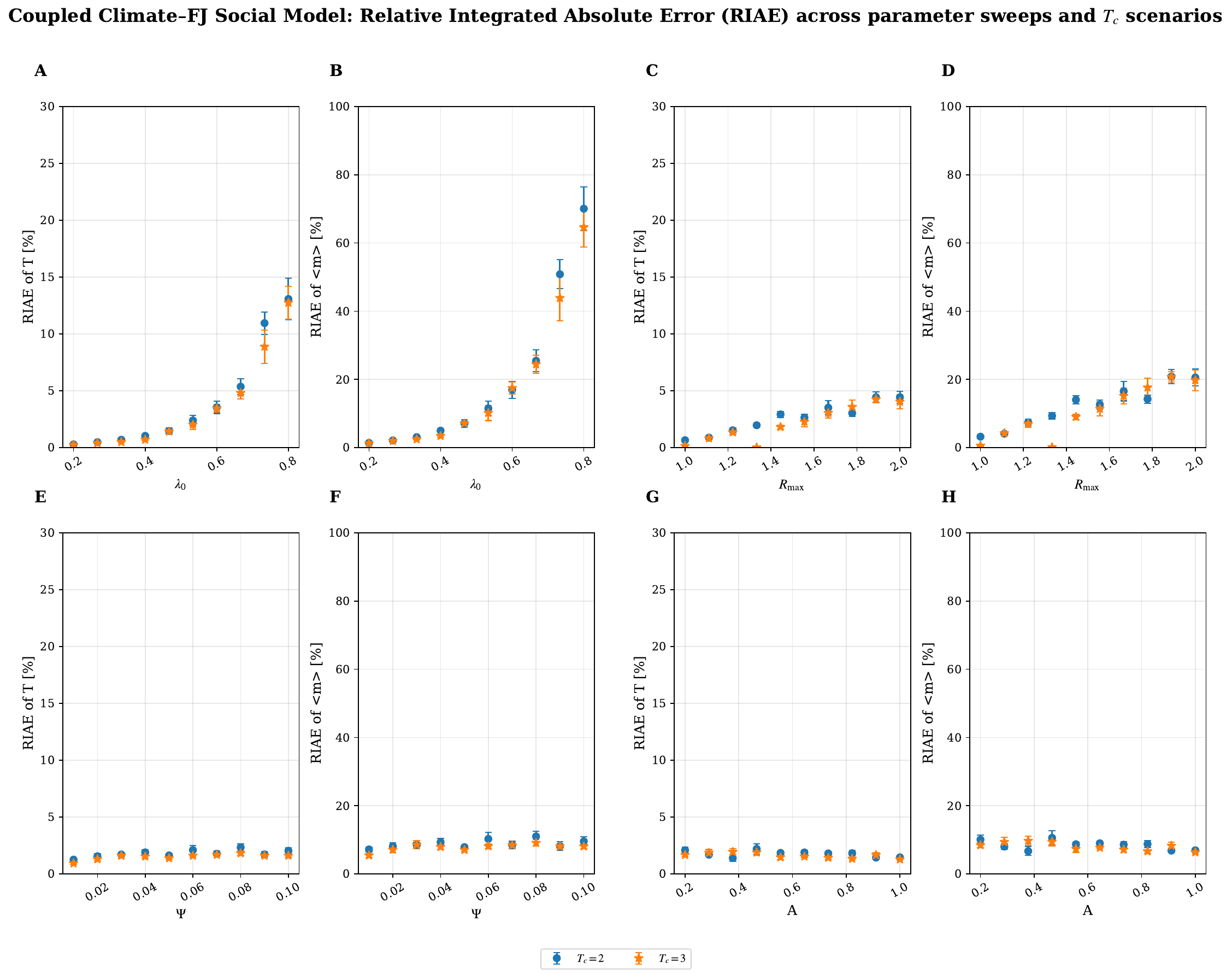}
\caption{%
Comparison of binary and spectrum FJ social models coupled to the climate–carbon system. 
Panels show RIAE of $T(t)$ (left) and $\langle m(t)\rangle$ (right) across sweeps of mean susceptibility $\lambda_0$, runaway amplitude $R_{\max}$, adjustment rate $\Psi$, and interaction sensitivity $A$. 
}
\label{fig:binary_vs_spectrum2}
\end{figure}

\subsection{Coupling Between the FJ Social Model and the Climate--Carbon Model}

The coupling between the FJ social dynamics and the climate--carbon model is mediated by two feedbacks:
temperature anomaly influences opinion change through a runaway forcing term, and mean social opinion
feeds back to alter effective emissions. In the FJ framework the coupling to the climate system occurs through the runaway forcing term $R(T)$, defined as
\begin{equation}
R(T) \;=\; -\,R_0 \;+\; \frac{R_{\max}}{1 + \exp[-\alpha (T - T_{\mathrm{c}})]},
\end{equation}
where $R_0$ is a baseline offset, $R_{\max}$ is the saturation amplitude, $\alpha$ is the steepness, and
$T_{\mathrm{c}}$ is a threshold temperature. As $T$ increases beyond $T_{\mathrm{c}}$, the logistic
function grows, increasing the pressure for opinions to shift toward pro-mitigation stances.

The collective outcome of the FJ dynamics is summarized by the average opinion. In the spectrum
formulation this is:
\begin{equation}
\bar X_{\mathrm{spec}}(t) \;=\; \frac{1}{n}\sum_{i=1}^n X_i(t),
\end{equation}
whereas in the binary version only the sign of each opinion is retained,
\begin{equation}
\bar X_{\mathrm{bin}}(t) \;=\; \frac{1}{n}\sum_{i=1}^n \mathrm{sgn}\big(X_i(t)\big).
\end{equation}

To systematically compare the binary ($N\!=\!2$) and spectrum ($N\!=\!100$) FJ social models when coupled
to the same climate–carbon system, we carried out a set of controlled parameter sweeps with stochastic
susceptibility and repeated runs per configuration. All simulations were run on the calendar interval 1800–2200. The social layer was kept inactive up
to the year 2017 so that climate variables evolve without social feedback. From 2017 onward, the social
dynamics were activated, and the coupled system was simulated through 2200. All comparisons and
error integrals were restricted to the active social window (2017–2200). For each parameter setting, we generated paired trajectories under the continuous and binary formulations and compared the climate temperature anomaly $T(t)$ and the mean social signal $\langle m(t)\rangle$ and report the RIAE. We varied one parameter at a time over 10 evenly spaced values while holding the others fixed at baselines.
The parameter ranges explored were:
\begin{itemize}
    \item Mean susceptibility $\lambda_0$: 0.2 to 0.8,
    \item Runaway amplitude $R_{\max}$: 1.0 to 2.0,
    \item Adjustment rate $\Psi$: 0.01 to 0.10,
    \item Interaction sensitivity $A$: 0.2 to 1.0.
\end{itemize}

Baseline values when not swept were fixed at
$R_0=0.778,\; R_{\max}=1.37,\; \Psi=0.07,\; A=0.35,\; \alpha=5.7,\; \lambda_0=0.5$.
Two ecological scenarios were considered, with $T_c \in \{2.0,\,3.0\}$ representing
low and high collapse thresholds. These baseline values were inspired by the work in \cite{kumar2025opinion}. Individual susceptibilities $\lambda_i$ are drawn at the start of each run from a clipped normal distribution centered at the sweep mean $\lambda_0$ with variance $0.1$, and then held fixed during the simulation. For each parameter value and each $T_c$ scenario, we executed
$N_{\mathrm{reps}}=10$ independent realizations (new draws of $\lambda$) for both the spectrum and binary
formulations. We averaged the resulting RIAE values across the 10 runs and report the mean with error
bars equal to the standard error of the mean. Overall, the results depicted in Figure~\ref{fig:binary_vs_spectrum2} shows the deviations between the binary and spectrum
formulations remain small across the vast majority of sweep values; in nearly all cases the RIAE for both $T(t)$ and $\langle m(t)\rangle$ stays below $10\%$, 
indicating that the binary model closely tracks the spectrum model under wide variation of 
susceptibility, runaway amplitude, adjustment rate, and interaction sensitivity. An exception 
occurs at the high extreme of $\lambda_0$, where the binary approximation begins to deviate 
more substantially from the spectrum formulation. Full information on simulation settings and 
the complete parameter table is provided in the Supplementary Information.

\subsection{Coupling of the Forest Model and the Replicator Social Model}

In this section we combine the forest dynamics with the replicator social dynamics to form a coupled
socio--ecological system. The coupling enters in two ways. First, individual utilities in the social model depend
on the ecological state: 
\[
U_i(F) \;=\; r_0 \,\big(2F-1\big)\,\Big( \frac{2i}{N-1} - 1 \Big),
\] 
so that the payoffs associated with mitigation categories vary with the fraction of forest cover $F$.  
Second, the replicator dynamics feed back into the ecological system through the mean social opinion
$\langle m \rangle = \sum_i m_i x_i$, where $m_i \in [-1,1]$ are fixed category opinions and $x_i$ are their
frequencies. This coupling enters the forest growth equation via a social feedback term
\[
J(x) \;=\; h \,\langle m \rangle,
\]
which shifts the balance of forest regeneration and loss depending on whether mitigation- or exploitation-oriented opinions dominate. To isolate the effect of social feedback, the simulations are run in two distinct phases. During the initial phase, the social model is switched off, allowing the forest fraction $F$ to evolve
autonomously until it reaches a quasi-equilibrium. At this point, the social dynamics are switched on, and the replicator system begins to co-evolve with the
forest state. From this switching point onward, we measure RIAE between the binary ($N=2$) and spectrum ($N=100$) replicator formulations for both the forest fraction $F$ and the mean opinion $\langle m\rangle$, continuing the simulation until the coupled system converges to a new equilibrium. Systematic parameter sweeps were performed across four key ecological and social parameters,  while holding all others fixed at baseline values 
($s=0.1$, $r_0=1.0$, $v=0.01$, $c=0.3$, $k=5.0$, $b=0.0$). 
For the baseline configuration, we adopted parameter values reported in \cite{innes2013impact}.
Specifically, we varied:
\begin{itemize}
    \item the forest loss rate $v$ from $0.0$ to $0.12$,
    \item the social learning rate $\kappa$ from $0.01$ to $0.10$,
    \item the ecological feedback strength $r_0$ from $0.1$ to $2.0$,
    \item the forest regeneration parameter $c$ from $0.1$ to $0.5$.
\end{itemize}
For each of these sweeps, we considered two values of the social--ecological coupling
$h \in \{0.01, 0.1\}$, representing weak versus strong influence of mean opinion on forest growth. Figure~\ref{fig:binary_vs_spectrum3} reports the results of these sweeps. Across most parameter 
ranges, the RIAE for both forest cover $F$ and mean opinion $\langle m\rangle$ remains low, generally 
below $10\%$, demonstrating that the binary approximation captures the spectrum dynamics with high 
fidelity. Slightly higher deviations can occur under extreme parameter values (e.g.\ low $r_0$ or high 
$h$), but overall the binary model tracks the spectrum formulation closely, even when forest and social dynamics are strongly coupled.

\begin{figure}[htbp]
\centering
\includegraphics[width=\textwidth]{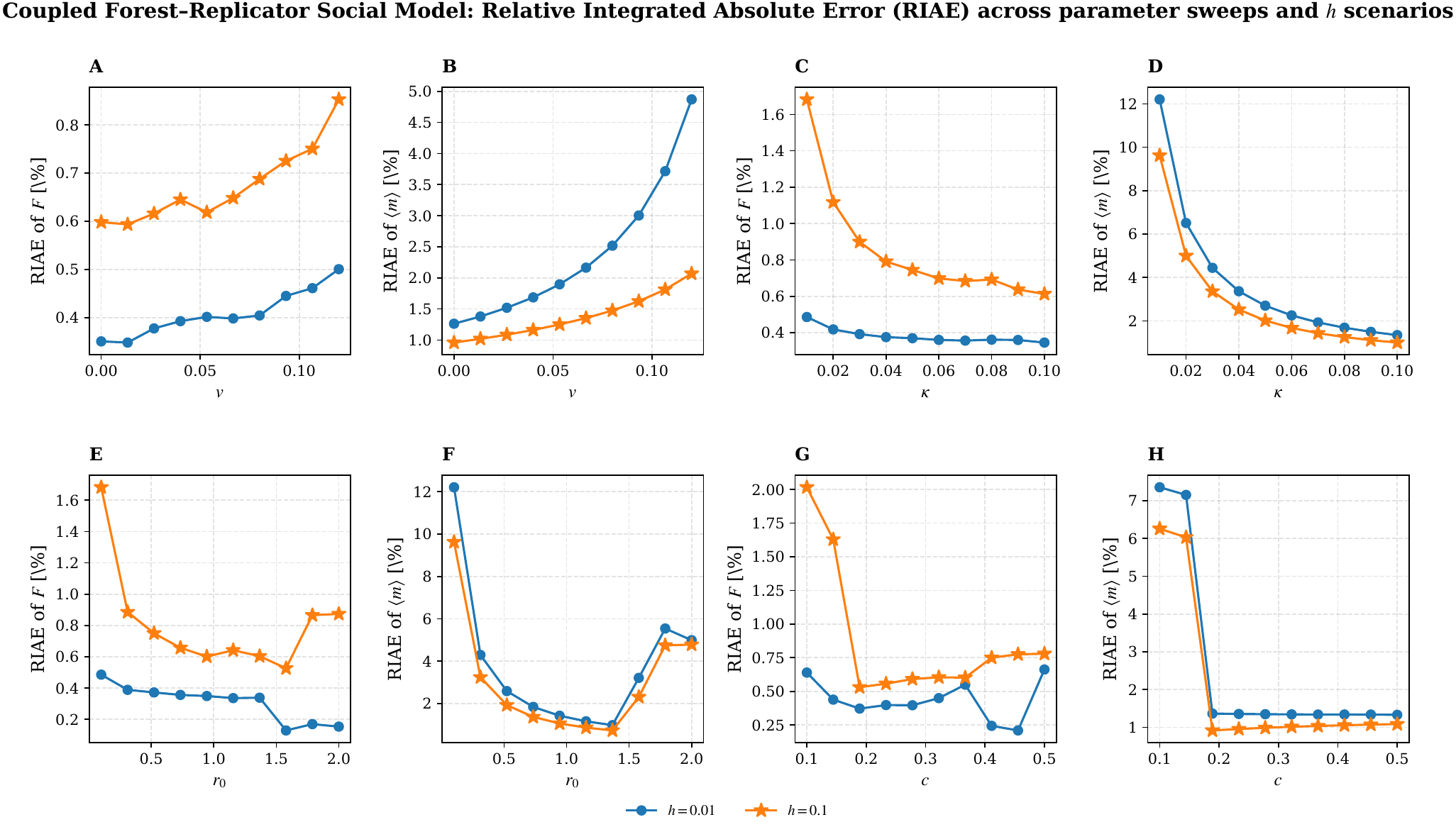}
\caption{%
Comparison of binary and spectrum replicator models coupled to the forest–grassland system. 
RIAE of forest fraction $F$ (left) and mean opinion $\langle m\rangle$ (right) are shown for sweeps of $v$, $\kappa$, $r_0$, and $c$, under weak ($h=0.01$) and strong ($h=0.1$) coupling. 
}
\label{fig:binary_vs_spectrum4}
\end{figure}

\subsection{Coupling of the Forest Model and the FJ Social Model}

The coupling between the forest dynamics and the FJ social dynamics is bidirectional and occurs through two feedback channels. First, the forest state $F$ influences social dynamics through the runaway feedback term,
\begin{equation}
R(T) \;=\; -\,R_0 \;+\; \frac{R_{\max}}{1 + \exp[-\alpha (F - F_{\mathrm{c}})]},
\end{equation}
which enters additively into the FJ opinion update. This term biases individuals toward more pro-ecological stances as the forest fraction approaches or falls below the critical threshold $F_c$.

Second, the mean social opinion feeds back into the forest through
\begin{equation}
J(x) \;=\; h\,\bar X(t),
\qquad
\bar X(t) = 
\begin{cases}
\frac{1}{N}\sum_{i=1}^N X_i(t), & \text{continuous case} \\[1ex]
\frac{1}{N}\sum_{i=1}^N \mathrm{sign}\!\big(X_i(t)\big), & \text{binary case}
\end{cases}
\end{equation}
which modifies the forest growth equation by adding or subtracting from the net regeneration rate depending on whether average opinion supports or resists conservation. As previously explained, the simulations are structured in two phases. In the first phase, the social dynamics are kept inactive, and the forest fraction $F$ evolves autonomously toward a quasi-equilibrium. In the second phase, the FJ social dynamics are activated and co-evolve with the forest state. From this switching point onward, we compute the relative integrated absolute error (RIAE) between the binary ($N=2$) and spectrum ($N=100$) FJ formulations for both the forest fraction $F$ and the mean opinion $\langle m\rangle$, continuing the simulation until the coupled system converges to a new equilibrium.

Unless being swept, parameters are fixed at
\[
R_0=1,\ \alpha=5,\ k=5,\ h=0.01,\ 
R_{\max}=1.5,\ \Psi=0.05,\ A=0.35,\ \lambda_0=0.5,\ c=0.3,\ v=0.01,
\]
 We examine two ecological threshold scenarios
\[
F_c \in \{0.2,\ 0.8\},
\]
representing a resilient versus fragile forest regime. We then vary, one at a time:
\begin{itemize}
    \item $R_{\max}$: 1.0 to 2.0,
    \item $\Psi$: 0.01 to 0.10,
    \item $A$: 0.2 to 1.0,
    \item $\lambda_0$: 0.2 to 0.8,
    \item $c$: 0.1 to 0.5,
    \item $v$: 0.0 to 0.12.
\end{itemize}
For each grid point we run 10 repeated realizations and report the mean RIAE with standard-error bars. Figure~\ref{fig:binary_vs_spectrum4} shows the outcomes of the parameter sweeps for the coupled Forest–FJ system. In nearly all cases, the deviation between the binary and spectrum models for the forest fraction $F$ remains minimal, with only larger discrepancies emerging under very low values of the turnover parameter $v$. For the social layer, the relative differences in mean opinion $\langle X\rangle$ are generally higher than for the forest dynamics, yet still remain within a range that provides a reliable approximation. Overall, these results demonstrate that the binary FJ formulation can capture the essential coupled dynamics with good fidelity across a wide spectrum of parameter regimes.

\begin{figure}[htbp]
\centering
\includegraphics[width=\textwidth]{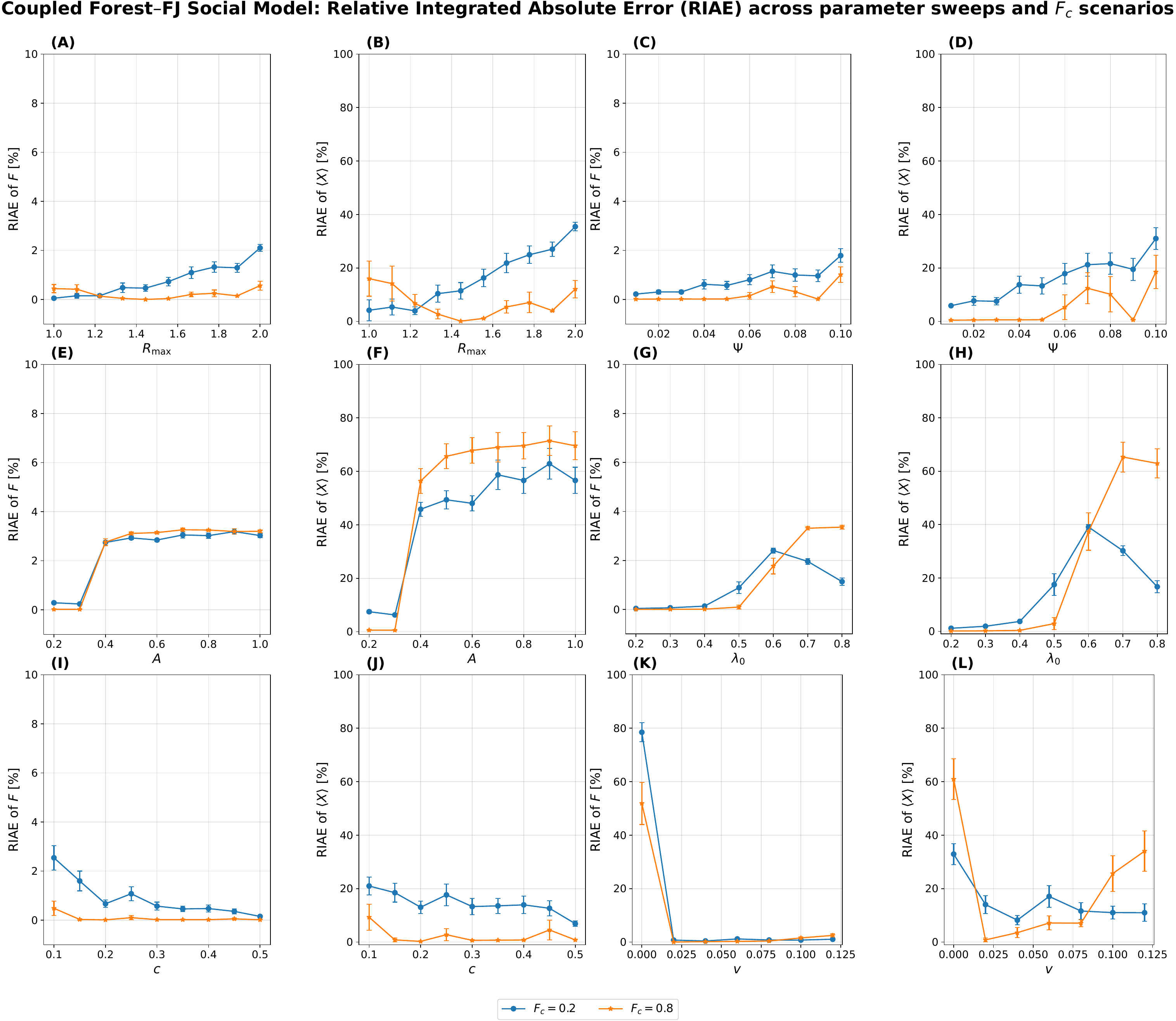}
\caption{%
Comparison of binary and spectrum FJ models coupled to the forest–grassland system. 
Panels report RIAE of forest fraction $F$ and mean opinion $\langle X\rangle$ across sweeps of $R_{\max}$, $\Psi$, $A$, $\lambda_0$, $c$, and $v$, under two ecological thresholds $F_c \in \{0.2,0.8\}$.
}
\label{fig:binary_vs_spectrum5}
\end{figure}

\section{Discussion}



In this work, our aim is not to claim that binary social models replicate the full behaviour of continuous formulations. Instead, we delineate the parameter regimes in which the binary approximation is adequate. Across the four couplings we examined, replicator vs.\ Friedkin--Johnsen dynamics paired with either the climate--carbon system or the forest--grassland mosaic, we identify broad regions where the binary model tracks its spectrum counterpart with errors typically within \(\sim 15\%\) RIAE, and often lower. Within these regimes, the salient feedbacks between human behaviour and ecological/climate processes are preserved well enough for comparative and policy-facing analyses.

This result should be read as qualified robustness, not equivalence. Continuous models retain descriptive richness and can deviate from binary behaviour outside the identified ranges or near sensitive thresholds. Nevertheless, by quantifying where the approximation holds, and its typical magnitude (RIAE \(\lesssim 15\%\) in most cases), we provide a practical guide for when the computational simplicity and interpretability of binary models can be leveraged without materially distorting system-level conclusions.

Despite these encouraging results, several limitations should be noted. First, our models necessarily involve
simplifications of both ecological and social processes. For instance, social dynamics are assumed to be homogeneous within categories and to follow fixed interaction rules, whereas in reality, social influence can be heterogeneous, adaptive, and context-dependent. Likewise, ecological models are stylized representations of climate and forest systems, omitting spatial heterogeneity, stochastic disturbances, and multi-species interactions. Second, our evaluation focuses on long-term equilibrium behavior and relative trajectory
agreement. More transient phenomena, such as short-term oscillations, abrupt cascades, or early-warning
signals, may reveal subtler differences between binary and spectrum formulations. Finally, the RIAE metric, though well-suited to capturing integrated deviations, does not directly assess potential differences in variance, distributional properties, or extreme outcomes.

Future work can address these limitations by incorporating richer forms of heterogeneity and complexity into both the social and ecological layers. For example, integrating network-based social structures, adaptive susceptibility, or multi-dimensional opinions could reveal conditions under which spectrum models diverge more strongly from binary approximations. On the ecological side, extending to spatially explicit landscapes or Earth system models of higher complexity would test whether binary models remain robust at larger scales. In addition, future studies could examine the performance of binary models in capturing leading indicators of critical transitions, such as flickering or rising variance, which are not the
focus of the present work. Finally, empirical validation through comparison with observed social and
ecological time series would help assess the practical utility of binary models for forecasting and policy design.

\section{Supplementary materials}

\subsection{Formulation of Climate–carbon model}

We have used the earth system model \cite{lenton2000land} combined with reduced ocean dynamics \cite{muryshev2015lag}. The full earth system model is as follows:

\begin{align}
    \frac{dC_{at}}{dt} &= \epsilon(t) - P + R_{veg} + R_{so} - F_{oc} \\
    \frac{dC_{oc}}{dt} &= F_{oc} \\
    \frac{dC_{veg}}{dt} &= P - R_{veg} - L \\
    \frac{dC_{so}}{dt} &= L - R_{so} \\
    c\frac{dT}{dt} &= (F_d - \sigma T^4)a_E
\end{align}

P represents photosynthesis which takes the following form:

\begin{equation}
    P(C_{at},T) = k_p C_{v0} k_{MM} (\frac{pCO_{2a}-k_c}{K_M + pCO_{2a}-k_c})(\frac{(15+T)^2 (25-T)}{5625}
\end{equation}

for $pCO_{2a}>= k_c$ and $-15 <= T <= 25$, and zero otherwise. $PCO_{2a}$ is defined as the ratio of moles of $CO_2$ in the atmosphere to the total number of moles of molecules in the atmosphere $k_a$:

\begin{equation}
    pCO_{2a}  = \frac{f_{gtm}(C_{at}+C_{at0})}{k_a}
\end{equation}

where $f_{gtm}=8.3259 * 10^{13}$ is the conversion factor from GtC to moles of carbon and $C_{at0}$ is the initial level of $CO_2$ in the atmosphere. 

Plant respiration takes the form:
\begin{equation}
    R_{veg}(T,C_{veg})  = k_r C_{veg} k_A e^{\frac{-E_a}{R(T+T_0)}}
\end{equation}

Soil respiration takes the form :
\begin{equation}
    R_{so}(T,C_{so})  = k_{sr} C_{so} k_B e^{\frac{-308.56}{T+T_0-227.13)}}
\end{equation}

Turnover (constant fraction of plants dying in a given unit of time) takes the form:
\begin{equation}
    L(C_{veg}) = k_t C_{veg}
\end{equation}

flux of $CO_2$ from the atmosphere to the ocean takes the form

\begin{equation}
    F_{oc}(C_{at},C_{oc}) = F_{0\chi}(C_{at} - \zeta \frac{C_{at0}}{C_{oc0} C_{oc}})
\end{equation}

where $\chi$ is the characteristic solubility of $CO_2$ in water and $\zeta$ is the evasion factor.

The net downward flux of absorbed radiation at the surface is :

\begin{equation}
    F_d = \frac{(1-A)S}{4}(1+ \frac{3\tau}{4})
\end{equation}

where A is the surface albedo, S is the incoming solar flux and $\tau$ is the vertical opacity of the greenhouse atmosphere. The opacity of each greenhouse is given by:

\begin{align}
    \tau(CO_2) &= 1.73(pCO_2)^{0.263} \\
    \tau(H_2O) &= 0.0126 (H P_0 e^{\frac{-L}{RT}})^{0.503} \\
    \tau(CH_4) &= 0.0231
\end{align}

H is the relative humidity, $P_0$ is the water vapor saturation constant, L is the latent heat per mole of water, and R is the molar gas constant.

\subsection{Full details of the replicator--climate--carbon model coupling}
\label{sec:replicator_climate_coupling}
 
The coupled system is solved numerically using the stiff solver \texttt{BDF} 
within \texttt{solve\_ivp}. The initial condition sets carbon pools to zero 
perturbation, and social shares are initialized such that on average only 
\(5\%\) of the population is mitigating at the start of the simulation. Parameter sweeps are performed over 
four key social parameters:
\[
\kappa,\quad \alpha_{\max},\quad \delta,\quad m_{\max},
\]
with both low and high critical temperature thresholds 
(\(T_c=2\) and \(T_c=3\)).

Table~\ref{tab:parameters} lists all constants and baseline values used in the 
simulation, divided into carbon--climate subsystem and social subsystem. 
Where parameters are varied in sweeps, the range is reported alongside the 
baseline.

\begin{table}[htbp]
\centering
\caption{Parameters of the replicator--climate--carbon coupled model.}
\label{tab:parameters}
\begin{tabular}{lll}
\hline
\textbf{Parameter} & \textbf{Value / Range} & \textbf{Description} \\
\hline
\multicolumn{3}{c}{\textit{Initial carbon pools}} \\
$C_{s0}$ & $1500$ & Soil carbon initial (GtC) \\
$C_{ve0}$ & $550$ & Vegetation carbon initial (GtC) \\
$C_{a0}$ & $596$ & Atmospheric carbon initial (GtC) \\
$C_{oc0}$ & $1.5 \times 10^{5}$ & Ocean carbon initial (GtC) \\
\hline
\multicolumn{3}{c}{\textit{Carbon--climate parameters}} \\
$f_{\mathrm{gtm}}$ & $8.3259 \times 10^{13}$ & Gas transfer constant \\
$K_a$ & $1.773 \times 10^{20}$ & Carbonate equilibrium constant \\
$K_p$ & $0.184$ & Photosynthesis prefactor \\
$K_A$ & $8.7039 \times 10^{9}$ & Vegetation growth factor \\
$K_{mm}$ & $1.478$ & Michaelis--Menten coefficient \\
$K_c$ & $29 \times 10^{-6}$ & Compensation concentration \\
$K_m$ & $120 \times 10^{-6}$ & Michaelis constant \\
$T_0$ & $288.15$ & Reference absolute temperature (K) \\
$K_r$ & $0.092$ & Vegetation respiration constant \\
$E_a$ & $54.83$ & Activation energy (kJ mol$^{-1}$) \\
$R$ & $8.314$ & Universal gas constant (J mol$^{-1}$ K$^{-1}$) \\
$K_{sr}$ & $0.034$ & Soil respiration constant \\
$K_b$ & $157.072$ & Soil carbon feedback factor \\
$K_t$ & $0.092$ & Vegetation litterfall rate \\
$F_0$ & $2.5 \times 10^{-2}$ & Ocean flux baseline \\
$\xi$ & $0.3$ & Ocean exchange coefficient \\
$\zeta$ & $50$ & Ocean buffer factor \\
$H$ & $0.5915$ & Humidity factor \\
$P_0$ & $1.4 \times 10^{11}$ & Reference pressure constant \\
Latent & $43655$ & Latent heat constant \\
$A$ & $0.225$ & Planetary albedo \\
$S$ & $1368$ & Solar constant (W m$^{-2}$) \\
$\text{capacity}$ & $4.69 \times 10^{23}$ & Heat capacity of climate system (J K$^{-1}$) \\
$\sigma$ & $5.67 \times 10^{-8}$ & Stefan--Boltzmann constant \\
$a_E$ & $5.101 \times 10^{14}$ & Effective emitting area (m$^2$) \\
$r_0$ & $5$ & Radiative forcing baseline factor \\
\hline
\multicolumn{3}{c}{\textit{Social subsystem parameters}} \\
$f_{\max}$ & $2.0$ & Maximum risk perception response \\
$\omega$ & $5.0$ & Steepness of logistic damage response \\
$T_c$ & $2.0,\,3.0$ & Critical temperature thresholds (varied) \\
$n_{\text{categories}}$ & $2$ (binary), $100$ (spectrum) & Opinion categories \\
$\kappa$ & Baseline $0.05$, range $[0.01,0.10]$ & Social learning rate \\
$\alpha_{\max}$ & Baseline $1.0$, range $[0.0,2.0]$ & Climate damage sensitivity \\
$\delta$ & Baseline $1.0$, range $[0.0,2.0]$ & Social reinforcement strength \\
$m_{\max}$ & Baseline $1.0$, range $[0.0,2.0]$ & Maximum mitigation utility \\

\hline
\end{tabular}
\end{table}

\subsection{Full details of the FJ social model--forest cover coupling}
\label{sec:fj_forest_coupling}

We first allow the forest subsystem to evolve in isolation. From \(F(0)=0.2\), we switch off social interactions and evolve only the forest equation up to \(t_{\mathrm{on}}=400\). This pre-activation phase allows the ecological state to relax toward its intrinsic trajectory without social pressure. At \(t_{\mathrm{on}}\) we switch on the FJ social layer and let opinions push back on the forest. To keep opinions physical, we confine \(X_i(t)\) to \([-1,1]\) by clipping states at every right-hand-side evaluation and zeroing any outward-pointing velocities at the bounds.

We explore two social initializations that mirror our binary vs.\ continuous comparison: a continuous mode (all \(X_i(0)=0\)) and a binary-proxy mode (half \(X_i(0)=-1\), half \(+1\), shuffled). After activation, we let the coupled system run until the forest response stabilizes; concretely, we declare convergence once
\(|\dot F(t)|<\texttt{tolF}=10^{-6}\) holds after a short buffer (\(t\ge t_{\mathrm{on}}+50\)) so that transient switching effects are excluded. Integration uses the stiff BDF solver with a uniform evaluation grid of \(\approx 50\) points per time unit, up to max time=$5000$.

For each parameter setting we simulate both modes, and quantify how much the two behaviors differ only in the post-activation window. We do so via phase-2 relative integrated absolute error (RIAE) metrics for forest cover \(F\) and the population mean \(\langle X\rangle\). We repeat this procedure across two runaway-threshold scenarios \(F_c\in\{0.2,0.8\}\) and perform one-dimensional sweeps over \(R_{\max},\,\Psi,\,A\), \(\lambda_0\), \(c\), and \(v\), keeping all other parameters at their baselines (Table~\ref{tab:fj_forest_params}). Each point aggregates 20 stochastic replicates (independent \(\Lambda_i\) draws with fixed seeds), and we report means with standard errors.

\begin{table}[h!]
\centering
\caption{Parameters and numerical settings for the FJ--forest coupled model. Baselines are used unless a sweep over the stated range is performed.}
\label{tab:fj_forest_params}
\begin{tabular}{lll}
\hline
\textbf{Symbol / Name} & \textbf{Baseline / Range} & \textbf{Role / Description} \\
\hline

\hline
\multicolumn{3}{c}{\textit{Forest-cover dynamics}} \\
$F(0)$ & $0.2$ & Initial forest cover \\

$c$ & Baseline $0.3$, sweep $[0.1,\,0.5]$ & Growth gain coefficient in $w(F)$ \\
$k$ & $5$ & Steepness inside $w(F)$ \\
$v$ & Baseline $0.01$, sweep $[0.0,\,0.12]$ & Linear loss rate \\
$h$ & $0.01$ (fixed) & Social push on $F$ via mean opinion \\
\hline
\multicolumn{3}{c}{\textit{Runaway / environmental drive}} \\

$R_0$ & $1$ & Baseline offset in $r_F$ \\
$R_{\max}$ & Baseline $1.5$, sweep $[1.0,\,2.0]$ & Max runaway pressure amplitude \\
$\alpha$ & $5$ & Runaway steepness \\
$F_c$ & $\{0.2,\,0.8\}$ & Threshold scenarios (blue/orange in plots) \\
\hline
\multicolumn{3}{c}{\textit{FJ social dynamics and heterogeneity}} \\
$\Psi$ & Baseline $0.05$, sweep $[0.01,\,0.10]$ & Social timescale (interaction strength) \\
$A$  & Baseline $0.35$, sweep $[0.2,\,1.0]$ & Interaction length in $W_{ij}=\exp(-|X_i-X_j|/A)$ \\
$\lambda_0$ & Baseline $0.5$, sweep $[0.2,\,0.8]$ & Mean susceptibility; $\lambda_i$ \\
$X_i^{(0)}$ & Mode 1: $0$; Mode 2: half $-1$, half $+1$ & Initial opinions (shuffled in mode 2) \\

\hline

\hline
\end{tabular}
\end{table}

\newpage
\subsection{Historical record of $CO_2$ emission}

\begin{figure}[htbp]
\centering
\includegraphics[width=0.8\textwidth]{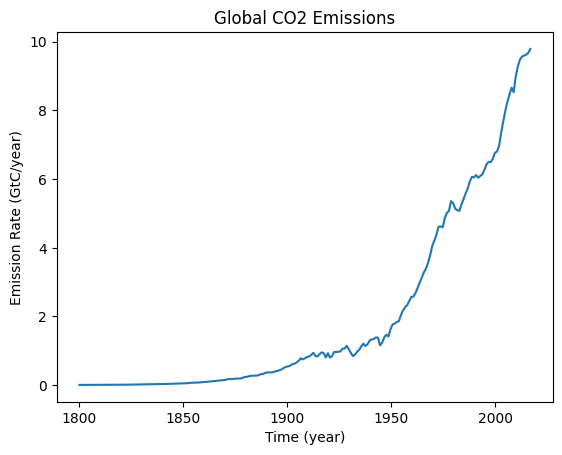} 

\caption{Historical record for Emission rate Vs Time is plotted. This data has been gathered from 1800 until 2017.}
\label{Mu Vs t}
\end{figure}

\newpage
\section{Data Availability}

All data and code supporting this study are available at: 
\newline
https://github.com/Yazdan-Babazadeh/Binary-Continuous-Social-Models

\newpage

\end{document}